%% file: paper.tex
\documentclass[conference]{IEEEtran}
\IEEEoverridecommandlockouts
\usepackage{cite}
\usepackage{amsmath,amssymb,amsfonts,amsthm}
\usepackage{stmaryrd}
\usepackage{algorithm}
\usepackage{algorithmicx}
\usepackage{algpseudocode}
\usepackage{graphicx}
\usepackage[caption=false]{subfig}
\graphicspath{{./figs/}}
\usepackage{textcomp}
\usepackage{xcolor}
\usepackage{xspace}
\usepackage{url}
\usepackage{thmtools,thm-restate}

\usepackage{mdframed}
\mdfsetup{%
  innertopmargin=4pt,
  innerbottommargin=4pt,
  innerleftmargin=4pt,
  innerrightmargin=4pt,
  leftmargin=-2pt,
  rightmargin=-2pt,
  backgroundcolor=blue!5,
  roundcorner=10pt
}
\usepackage{listings}
\definecolor{darkgreen}{rgb}{0.1, 0.5, 0.3}
\usepackage{ifthen}
\usepackage{accents}
\usepackage{dsfont}
\usepackage{siunitx}
\usepackage{amsthm}
\usepackage{empheq}
\usepackage{multirow}
\usepackage{bigdelim}
\usepackage{tabularx}
\usepackage{xfrac}
\DeclareMathOperator*{\argmin}{arg\,min}
\DeclareMathOperator*{\argmax}{arg\,max}
\newcolumntype{s}{>{\scriptstyle}c}
\newcolumntype{Y}{>{\centering\arraybackslash}X}
\newcommand\rvb[4]{{\multirow{#1}{*}{$\hspace{#2pt}\left.\rule{0pt}{#3pt}\right\}{#4}$}}}
\newcommand\smo{{\!{\scriptscriptstyle-}1\!}}
\newcommand\bz{{\hspace{-1pt}\bf0\hspace{-0.8pt}}}
\newcommand{\Rpos}{{\mathbb{R}_{\scriptscriptstyle+}}}
\newcommand{\Rneg}{{\mathbb{R}_{\scriptscriptstyle-}^\ast}}
\newcommand{\binset}{{\mathbb{Z}_2}}
\newcommand{\norm}[2]{\|#2\|_{#1}}
\newcommand{\new}[1]{{\widetilde{#1}}}
\newcommand{\err}{\vec{e}}

\newcommand{\gemma}{\textsf{Gemma}\xspace}

\newcommand{\model}{CCM\xspace}
\newcommand{\dlb}{CCM-LB\xspace}
\newcommand{\code}[1]{\textsf{#1}}
\newcommand{\taskset}[2]{{T_{#2}^{#1}}}
\newcommand{\loadsym}{{\mathcal{L}}}
\newcommand{\load}[2]{{\loadsym^{#1}(#2)}}
\newcommand{\newload}[2]{{\new{\loadsym}^{#1}(#2)}}
\newcommand{\sharset}[2]{{S_{#2}^{#1}}}
\newcommand{\newsharset}[2]{{\new{S}_{#2}^{#1}}}
\newcommand{\homeset}[2]{{\widehat{S}_{#2}^{#1}}}
\newcommand{\memory}[3]{{\mathcal{M}_{#1}^{#2}\!\left(#3\right)}}
\newcommand{\newmemory}[3]{{\new{\mathcal{M}}_{#1}^{#2}(#3)}}
\newcommand{\normmem}[2]{{\memory{}{#1}{#2}}}
\newcommand{\basemem}[2]{{\memory{\scriptscriptstyle-}{#1}{#2}}}
\newcommand{\overmem}[2]{{\memory{\scriptscriptstyle+}{#1}{#2}}}

\newcommand{\maximem}[2]{{\memory{\mathrm{\max}}{#1}{#2}}}
\newcommand{\taskmem}[2]{{\memory{T}{#1}{#2}}}
\newcommand{\sharmem}[2]{{\memory{S}{#1}{#2}}}
\newcommand{\homemem}[2]{{\memory{H}{#1}{#2}}}
\newcommand{\newhomemem}[2]{{\newmemory{H}{#1}{#2}}}
\newcommand{\avaimem}[1]{{\memory{\scriptscriptstyle\infty}{}{#1}}}
\newcommand{\recv}{{\gets}}
\newcommand{\sent}{{\to}}
\newcommand{\commset}[2]{{C_{#2}^{#1}}}
\newcommand{\recvset}[2]{{\accentset{\recv}{\commset{#1}{#2}}}}
\newcommand{\sentset}[2]{{\accentset{\sent}{\commset{#1}{#2}}}}
\newcommand{\volume}[4][]{{\mathcal{V}_{#2}^{#3}(#4\ifthenelse{\equal{#1}{}}{}{,#1})}}
\newcommand{\volon}[2]{{\volume{}{#1}{#2}}}
\newcommand{\voloff}[2]{{\volume{\scriptscriptstyle\notin}{#1}{#2}}}
\newcommand{\work}[2][]{{\mathcal{W}_{#1}^{#2}}}
\newcommand{\workmod}[2]{{\work{#1}(#2)}}
\newcommand{\workmax}[1]{{\work[\max]{#1}}}
\newcommand{\sbt}{u}
\newcommand{\sbh}{v}
\newcommand{\ctt}{w}
\newcommand{\transp}[1]{{{#1}^\intercal}}

\newcommand{\ii}[1]{{\llbracket1,{#1}\rrbracket}}
\newcommand{\forallint}[2]{{\left(\forall{#1}\in\ii{#2}\right)}}
\newcommand{\slc}[3]{{({#1}^{#2}_{::{#3}})}}
\newcommand{\sss}[3]{$\scriptscriptstyle{#1}^{#2}_{::{#3}}$}
\newcommand{\upen}[2][]{{\mathcal{U}_{#2}\ifthenelse{\equal{#1}{}}{}{(#1)}}}

\algnewcommand\algorithmicforeach{\textbf{foreach}}
\algdef{S}[FOR]{ForEach}[1]{\algorithmicforeach\ #1\ \algorithmicdo}

\begin{document}

\title{A Communication- and Memory-Aware Model\\for Load Balancing Tasks
\thanks{
    Sandia National Laboratories is a multimission laboratory managed and
    operated by National Technology \& Engineering Solutions of Sandia, LLC, a
    wholly owned subsidiary of Honeywell International Inc., for the
    U.S. Department of Energy’s National Nuclear Security Administration under
    contract DE-NA0003525. This paper describes objective technical results and
    analysis. Any subjective views or opinions that might be expressed in the
    paper do not necessarily represent the views of the U.S. Department of
    Energy or the United States Government. ~ cf. \texttt{SAND2024-05173C}
}
}
\author{
  Jonathan Lifflander$^\ast$,
  Philippe P. P\'eba\"y$^\dagger$,
  Nicole L. Slattengren$^\ast$,\\
  Pierre L. P\'eba\"y$^{\dagger\diamond}$,
  Robert A. Pfeiffer$^\ast$,
  Joseph D. Kotulski$^\ast$,
  Sean T. McGovern$^\dagger$\\[2pt]
\IEEEauthorblockA{
$^\ast$\textit{Sandia National Laboratories}, Livermore, CA, U.S.A.}
\IEEEauthorblockA{
$^\dagger$\textit{NexGen Analytics}, Sheridan, WY, U.S.A.}
\IEEEauthorblockA{
$^\diamond$\textit{ISIMA School of Engineering}, Aubi\`ere, France}

}

\maketitle

\begin{abstract}
\input{abstract}

\end{abstract}


\input{10}
\input{20}

\input{30}

\input{40}
\input{50}
\input{70}

\input{80}

\clearpage

\bibliographystyle{IEEEtran}
\bibliography{bib/cited,bib/contact,bib/group,bib/this}

\clearpage

\appendices

\input{100}
\input{90}

\end{document}

%% file: abstract.tex
While load balancing in distributed-memory computing has been well-studied, we
present an innovative approach to this problem: a unified, reduced-order model
that combines three key components to describe ``work'' in a distributed system:
computation, communication, and memory.
Our model enables an optimizer to explore complex tradeoffs in task
placement, such as increased parallelism at the expense of data
replication, which increases memory usage.
We propose a fully distributed, heuristic-based load balancing
optimization algorithm, and demonstrate that it quickly finds
close-to-optimal solutions.
We formalize the complex optimization problem as a mixed-integer
linear program, and compare it to our strategy.
Finally, we show that when applied to an electromagnetics code, our
approach obtains up to 2.3x speedups for the imbalanced execution.

%% file: 10.tex
\section{Introduction}
\label{sec:intro}

As Moore's law has arguably ended and the exascale era emerges,
scientific applications are expected to run at larger scales to
decrease time-to-solution.
However, distributed-memory architectures have become more challenging
to program efficiently.
Achieving optimal performance often requires careful coordination and
mapping of data along with computational work to the available
hardware resources.
Developers are often forced to make difficult decisions in trading off
parallelism for communication, data replication, and memory use that
may not be portable across different platforms.
Task-based programming models have emerged as a possible solution, especially
for irregular computational structures where manual work partitioning is
particularly challenging.
Instead of decomposing a problem to a fixed number of MPI ranks at
startup, the programmer exposes concurrency to a middleware runtime
system in the form of migratable tasks that can execute on different
and possibly heterogeneous compute nodes.

Task-based paradigms vary greatly on the level and detail of
information passed to the middleware, and the user is still often
expected to make good decisions in breaking down work into tasks to
get optimal performance.
However, for a tasking model to be performance portable, where each
task should run to get optimal performance (i.e., \emph{load balancing})
is a key problem that should be passed to the middleware.
For load balancing to be automated by the middleware, the profile of
tasks must be predictable to some extent, or an online balancing scheme
(e.g., work stealing) must be used. 
Although making online schemes locality-aware has been studied (cf.~\S\ref{sec:related}),
these approaches still have many limitations in achieving optimal
performance, especially under tight memory constraints.

This article thus makes the fundamental assumption that task profiles can be
predicted by using a \emph{cost model}.  We use these predictions to propose a
novel work model, called \model (\underline{C}omputation
\underline{C}ommunication \underline{M}emory), to describe the amount of work
that each processor is performing under a task-to-processor mapping.  Its
primary contributions include:
\begin{itemize}
\item formulating the load balancing problem into a model that can be used to
  trade-off communication, processor load, and data replication under memory
  constraints;
\item a fully distributed load balancing algorithm (called \dlb) that
  uses the \model model to redistribute tasks;
\item a recasting as a mixed-integer linear program (MILP) of the
  \model optimization problem, to validate that \dlb finds solutions 
  at worst 1.8\% slower than optimal ones;
\item a machine learning approach for the non-iterative (i.e., without
   repetitive behaviour across iterations) target application to
   predict task durations fed into \dlb; and,
\item an application of our approach to an electromagnetics code,
  demonstrating a 2.3x speedup for the imbalanced matrix assembly on
  128 nodes.
\end{itemize}

%% file: 20.tex
\section{Related Work}
\label{sec:related}
Load balancing is a well-known and extensively studied problem.
Regularly-structured applications often achieve load balance through
data distribution and careful orchestration of communication (e.g.,
multipartitioning~\cite{DBLP:journals/jpdc/DarteMFC03}).
For subclasses of
regularly-structured computations (e.g., affine loops), extensive research has
studied the memory and communication tradeoffs for generating distributed-memory
mappings~\cite{reddy2014effective,bondhugula2013compiling}. This has been
extended to broader classes such as tensor
computations~\cite{kong2023automatic}.
Inspector-executor approaches can often apply to calculations involving sparse
matrices~\cite{DBLP:journals/tpds/CatalyurekA99} or
meshes~\cite{Williams:1991:PDL:124737.124739} where computational data and load
can be analyzed \emph{a priori} at runtime before it starts (e.g.,
CHAOS~\cite{parti}).
Partitioning schemes~\cite{Hendrickson:1995:ISG:203046.203060}, whose
parallelization~\cite{DBLP:journals/jpdc/CatalyurekBDBHR09,karypis1997parmetis,zoltan2000}
is non-trivial, are often applied in this context.
However, the data replication within memory constraints, which graph
partitioners typically do not consider, must be explored for our
target application.

Online dynamic load balancing approaches such as \textsf{Cilk}'s work
stealing~\cite{Blumofe95} are widely studied, with provably optimal
space and time bounds for uniform shared-memory machines on
fully-strict parallelism.
Such schedulers have been extended for distributed 
memory~\cite{dinan-sc09}, but ignore data locality and memory
limits.
Subsequent work in shared-memory has extended work stealing to be
more aware of data locality, such as in hierarchical place
trees~\cite{yan2010hierarchical} and similar
approaches~\cite{acar-spaa00}. %

Task mapping has been studied for dataflow runtimes (e.g.,
PaRSEC~\cite{bosilca2013parsec}, StarPU~\cite{AugThiNamWac10CCPE},
Legion~\cite{bauer2012legion}), but often requires the user to generate an
efficient mapping. To express an application with dataflow often requires a
complete rewrite with data use types exposed (which is impractical for many real
applications), and exposes many degrees of freedom to explore. Recent
work~\cite{sfx2023automated} has shown that automated mappers may be feasible
but difficult to scale.
A MILP-based approach for mapping tasks to hardware is given
by~\cite{huang:2018}, but the description is terse and it is not
immediately clear how the Boolean constraints are converted into
integer ones, which is necessary if they are to be resolved by a MILP solver.

For iterative applications, scalable persistence-based load balancers that are
hierarchical~\cite{HierLdbIJHPCA10} or distributed~\cite{menon:13} may be
applied. However, these strategies often do not consider communication and lack
the ability to consider data replication and memory constraints.

Various parallel computation models~\cite{zhang2007models} have been proposed to
model the costs of executing a parallel program on hardware, such as
PRAM~\cite{aggarwal1990communication} and LogP~\cite{o:1}. These models are in
the same vein as our proposed model, \model.

\subsection{Background \& Challenges}
\label{sec:load_balancing_algorithm}
In this article, we devise and build a novel work model combining
three elements:
(1) computation (time spent executing a task),
(2) communication between tasks, and
(3) memory utilized by tasks (including shared memory).
The complex interplay between these elements creates an combinatorially
large search space for finding the optimal task assignments across nodes.
Furthermore, we propose a scalable algorithm to efficiently
search this space in an incremental manner, so task assignments can be refined
over time. Such a tunable algorithm should allow users to trade off quality with
time complexity, and thus lower the cost of running the load balancer.

The goal of a load balancer is to minimize the total time an application spends
working. A corollary to this is minimizing the longest time any rank spends
working. Thus, a load balancer may try to reduce the highest rank load,
$\max\loadsym$, to be as close as possible to the population mean of loads
across all ranks, $\mu_\loadsym$. A simple statistic to assess \emph{load
imbalance} is $\mathcal{I}_\loadsym:=\sfrac{\max\loadsym}{\mu_\loadsym}-1$,
vanishing if and only if all ranks have the same load, so none of them slow down
the rest. However, while minimizing imbalance is necessary to obtain an optimal
configuration, it is not sufficient if the total amount of work can vary. For
instance, displacing tasks from one rank to another may result in more overall
communication across slower off-node network edges, thereby increasing total
work and resulting in a longer execution time. Thus, a load balancing
algorithm must rather minimize the total work across all ranks while also
minimizing the maximum work performed on any rank.

The scalability of an application is further limited by the scalability of the
load balancer itself.
For large scales, fully distributed load balancing schemes show the
most promise. However, the quality of the distributions produced and the
complexity of implementation have traditionally limited their efficacy in
practice.  Of particular interest to us have been fully-distributed
\emph{epidemic} (or \emph{gossip}) algorithms, that distribute information
across the ranks to be rebalanced, in a manner similar to that of an infectious
disease spreading through a biological population.  This approach has shown
promise in an array of distributed applications, ranging from routing
protocols~\cite{vahdat2000epidemic} to database
consistency~\cite{demers1987epidemic}.
By building on original gossip-based work by Menon, et
al.~\cite{menon:13}, we propose a novel distributed load
balancing algorithm that optimizes task placement to minimize work while
operating under strict memory constraints.

%% file: 30.tex
\section{Definitions \& Models}
\label{sec:defs}
We start by specifying terminology whose meaning varies throughout
the literature, before introducing new concepts and
mathematical formulations of importance to our approach.
\subsection{Parallel Model}
\subsubsection{Nodes \& Ranks}
\label{sec:ranks}
A \emph{node} is the smallest compute unit connected to the network.
A \emph{rank} is a distinct process with a dedicated set of resources
(e.g., CPU cores, GPUs) belonging to a node.
The set of all ranks in the computation is denoted~$R$.
\subsubsection{Phases}
\label{sec:phases}
This paper focuses on load balancing a \emph{phase}: a set of tasks across
ranks that are to be executed between two synchronization points. For some
scientific applications, a phase may be an iteration or timestep that evolves
over time. For others, it may be the entire application's execution. The
proposed approach assumes that the tasks and communications are known or at
least can be predicted (either by modeling, persistence, or an
inspector-executor approach).
\subsubsection{Tasks}
\label{sec:tasks}
We define a \emph{task} as a potentially multi-threaded, non-preemptable sequence
of instructions that has a set of inputs and outputs, which include
\emph{communications}.  Each \emph{task} has an associated context in which it
executes, consuming memory, and it may produce outputs that can subsequently
spawn other tasks in other contexts.  The set of tasks present during a
phase~$p$ is denoted~$\taskset{p}{}$ and the set of tasks on rank~$r$ during
phase~$p$ is denoted~$\taskset{p}{r}$.
\subsubsection{Shared memory blocks}
\label{sec:shared_blocs}
These are memory chunks accessed by
multiple tasks, to either read them or perform commutative and associative
update operations on them.
A shared block~$s$  has a set of tasks $\taskset{p}{s}$ accessing it,
and it may be replicated across ranks to increase parallelism at the
cost of higher communication and memory use.
Each task $t$ is thus associated with a set of shared blocks
$\sharset{p}{t}$; each rank~$r$ is associated  with
$\sharset{p}{r}$$:=$$\cup_{t\in\taskset{p}{r}}\sharset{p}{t}$; and,
$\sharset{p}{}$$:=$$\cup_{r\in{R}}\sharset{p}{r}$ is the set of all shared
blocks.
To limit complexity, we assume that each task accesses at most one
shared block, so that $\sharset{p}{t}$ is either $\varnothing$ or a 
singleton\footnote{Access to multiple shared blocks does not
substantially alter the mathematical formulation, but impacts the
algorithmic treatment presented thereafter.}.

\subsection{Compute Model}
\label{sec:load_model}
A \emph{compute model} is an abstraction that predicts the required
time (in \SI{}{\second}) to complete the computations contained in
task~$t$ at phase~$p$, denoted~$\load{p}{t}$.
Denoting $\taskset{p}{r}$ the set of tasks present on rank~$r$ at
phase~$p$, the load of~$r$ is readily computed as:
\begin{equation}
\label{eqn:load_model}
\load{p}{r} := \textstyle\sum_{t\in\taskset{p}{r}}\load{p}{t}.
\end{equation}
Evidently, it is not necessary to recompute rank loads
using~\eqref{eqn:load_model} when transferring a task~$t$ from a
rank~$r_1$ to another rank~$r_2$; the following \emph{update formul\ae}
can be used instead:
\begin{equation}
\label{eqn:update_load_model}
\newload{p}{r_1} = \load{p}{r_1} - \load{p}{t},\quad
\newload{p}{r_2} = \load{p}{r_2} + \load{p}{t}.
\end{equation}

\subsection{Communication Model}
\label{sec:communication_model}
The set of input and output communications of a task~$t$ at phase~$p$
are respectively denoted $\recvset{p}{t}$ and $\sentset{p}{t}$, and
\begin{equation}
C^p:=
\bigcup_{\;r\in{}R^{\phantom{p}}_{\phantom{r}}}\!\bigcup_{t\in\taskset{p}{r}}\recvset{p}{t}
=\bigcup_{\;r\in{}R^{\phantom{p}}_{\phantom{r}}}\!\bigcup_{t\in\taskset{p}{r}}\sentset{p}{t},
\end{equation}
thanks to the symmetry between inter-task inputs and outputs.
In other words, the across-rank, across-task union of sent
communications is equal to that of received ones.
The \emph{volume} of communications sent from task~$t_1$ and received
by task~$t_2$ during phase~$p$ (measured in bytes (\SI{}{\byte})) is
denoted~$\volume[t_2]{}{p}{t_1}$.
Inter-task communications are aggregated at the rank level, as
\begin{equation}
\label{eqn:rank_agg_comm}
\volume[r_2]{}{p}{r_1} :=
\textstyle\sum_{(t_1,t_2)\in\taskset{p}{r_1}\times\taskset{p}{r_2}}
\volume[t_2]{}{p}{t_1}.
\end{equation}
In particular, $\volon{p}{r}:=$$\volume[r]{}{p}{r}$ is the
total on-rank communication volume for~$r$, whose time cost per byte
is orders of magnitude smaller than off-rank time cost per byte, defined
as\footnote{Our model assumes that incoming and outgoing
communications occur concurrently, hence we take the maximum between those.}:
\begin{equation}
\label{eqn:comm_model_off}
\voloff{p}{r} := \max\Biggl(%
\hspace{-4ex}\sum_{\hspace{2em}r_0\in R\setminus\{r\}}\hspace{-2em}\volume[r_0]{}{p}{r},
\hspace{-4ex}\sum_{\hspace{2em}r_0\in R\setminus\{r\}}\hspace{-2em}\volume[r]{}{p}{r_0}\Biggr).
\end{equation}
\subsection{Memory Model}
\label{sec:memory_model}
Each node has a fixed amount of random-access memory, and thus
the load balancer must prescribe a \emph{feasible} task
redistribution, so as not to exceed this memory limit.
Being mapped to a certain node, each rank~$r$ thus partakes of this
limit, for it has a baseline working memory usage $\basemem{p}{r}$
measured at the start of phase $p$, including base process usage and
application data structures.
Moreover, each task $t$ itself has baseline memory
usage~$\basemem{p}{t}$, always used, and overhead working memory
$\overmem{p}{t}$ during execution\footnote{The
task execution model is non-preemptable; thus, only one task will ever
be executed at once.}, neither of which includes memory for any shared blocks it might access.
These task memory components are then assembled at the rank level:
\begin{equation}
\label{eqn:taskmem}
\taskmem{p}{r} := \sum_{t\in\taskset{p}{r}}\!\basemem{p}{t}
+ \max_{t\in\taskset{p}{r}}\overmem{p}{t}.
\end{equation}
Furthermore, each shared block $s$ has a maximum amount of working
memory it may consume during phase $p$, denoted $\normmem{p}{s}$.
The size of the shared blocks operated on~$r$ is thus
$\sharmem{p}{r}:=\sum_{s\in\sharset{p}{r}}\!\normmem{p}{s}$.
The \emph{maximum memory usage} combines the baseline, task, and
shared memory components:
\begin{equation}
\label{eqn:rank_max_memory_usage}
\maximem{p}{r} := \basemem{p}{r} + \taskmem{p}{r} + \sharmem{p}{r}.
\end{equation}
Consequently, if $\avaimem{\mathfrak{n}}$ is the \emph{available
memory} on a given node~$\mathfrak{n}$, i.e., the upper limit on the
combined memory usage for all ranks on~$\mathfrak{n}$, the following
constraint must hold:
\begin{equation}
\label{eqn:memory_ub_node}
\sum_{r\in\mathfrak{n}}\maximem{p}{r}\leq\avaimem{\mathfrak{n}}.
\end{equation}
To further reduce complexity, we apply the more
stringent per-rank memory condition:
\begin{equation}
\label{eqn:memory_ub}
(\forall{}r\!\in\mathfrak{n})\enskip\maximem{p}{r}\leq\avaimem{r}
:= \tfrac{1}{\vert\{r_0\in\mathfrak{n}\}\vert}\avaimem{\mathfrak{n}},
\end{equation}
which is evidently sufficient for~\eqref{eqn:memory_ub_node}, but not
necessary to it.

The \emph{home} of a shared block that will be read by tasks is the rank
on which it is initialized.
For shared blocks that are being updated, however, the home is
uniquely defined as the rank on which the fully computed shared block
will be consumed in a subsequent phase, which is typically the rank on
which the tasks that update it were initialized.
The set of all shared blocks homed at rank~$r$ for phase~$p$ is
denoted~$\homeset{p}{r}$.
A shared block requires extra communication when computed on or read
from any rank other than its home, a cost which in our model is imputed
to the rank with the off-home shared block, as:
\begin{equation}
\label{eqn:homing_model}
\homemem{p}{r} := \textstyle
\sum_{s\in\sharset{p}{r}\setminus\homeset{p}{r}}\normmem{p}{s}.
\end{equation}
In a manner similar to what we did for load
in~\eqref{eqn:update_load_model}, we derive update formul\ae{} for
homing costs, when a task $t\in\taskset{p}{r_1}$ is
transferred from a rank~$r_1$ to another rank~$r_2$, resulting in new
shared block sets $\newsharset{p}{r_1}$ and $\newsharset{p}{r_2}$:
\begin{restatable}[Homing communications update formul{\ae}]{theorem}{thmhomingupdate}
\hfill\\[-8pt]
\begin{align}
\label{eqn:memory_model_sub_1}
\newhomemem{p}{r_1} &= \homemem{p}{r_1} - \textstyle
\sum_{s\in(\sharset{p}{t}\setminus\newsharset{p}{r_1})\setminus\homeset{p}{r_1}}
\normmem{p}{s},\\
\label{eqn:memory_model_sub_2}
\newhomemem{p}{r_2} &= \homemem{p}{r_2} + \textstyle
\sum_{s\in(\sharset{p}{t}\setminus\sharset{p}{r_2})\setminus\homeset{p}{r_2}}
\normmem{p}{s}.
\end{align}
\end{restatable}
\begin{proof}
Omitted for brevity, refer to Appendix~\ref{sec:proofs}.
\end{proof}
\subsection{\model Model}
\label{sec:work_model}
Our proposed \model model incorporates all components defined
separately above in~\eqref{eqn:load_model}, \eqref{eqn:rank_agg_comm},
\eqref{eqn:comm_model_off}, and \eqref{eqn:homing_model}, and
the memory constraint~\eqref{eqn:memory_ub} in the form of a
possibly-infinite penalization term, within the following
quasi\footnote{because the~$\varepsilon$ term is neither constant nor
linear constant.}-affine combination:
\begin{equation}
\label{eqn:work_model}
\workmod{p}{r} := \alpha\load{p}{r} + \beta\voloff{p}{r}
 + \gamma\volon{p}{r} + \delta\homemem{p}{r} + \varepsilon,
\end{equation}
with the following coefficients:
\begin{center}
\setlength{\arrayrulewidth}{1pt}
\begin{tabular}{@{}llll@{}}\hline
coefficient & unit & support & description\\\hline
$\alpha$ & $\varnothing$ & $\binset$
&exclusion/inclusion of~\eqref{eqn:load_model}\\
$\beta, \gamma$ & \SI{}{\second/\byte} & $\Rpos$
&scales \eqref{eqn:rank_agg_comm} \& \eqref{eqn:comm_model_off} to time\\
$\delta$ & \SI{}{\second/\byte} & $\Rpos$
&scales \eqref{eqn:homing_model} to time\\
$\varepsilon$ & \SI{}{\second} & \small$\{0;+\infty\}$
&{\small0} if \eqref{eqn:memory_ub} holds, else {\small$+\infty$}\\\hline
\end{tabular}
\end{center}
The scaling coefficients $\beta$, $\gamma$, and $\delta$ may
be measured empirically on a per-system basis, or 
evaluated from first principles.

In order to update the $\beta$ and $\gamma$ terms in the work model,
the intra- and inter-rank communication must also be updated.
This is done by finding edges that change from local to remote when
a task is transferred between ranks.
For brevity, the update formul\ae{} are not presented here, but
can be derived from
equations~\eqref{eqn:rank_agg_comm}, and~\eqref{eqn:comm_model_off}.

%% file: 40.tex
\section{Distributed \& Constrained Load Balancing}
\label{sec:mem-constrained-lb}
We now present our distributed algorithm, \dlb, which iteratively
optimizes task placement to reduce overall work as computed by the
\model model.
First, each iteration builds a distributed peer network for each rank
by propagating rank-local information.
Second, ranks can attempt to transfer work within their known peer
set, by evaluating a criterion using only locally-known information.

\begin{figure}
\input{ccm-lb-segment-1}
\caption{The \dlb algorithm.}
\label{alg:ccm_lb}
\end{figure}

Before \dlb builds peer networks, we first generate clusters of tasks (on each
rank) that are highly connected based on the weights of the \model model. These
clusters are generated based on tasks that communicate heavily or access the
same shared memory block(s). Clusters are important to consider migrating
together as splitting them apart often increases the amount of work in the
system or increases memory usage (since the same shared block will be accessed
by more ranks as they are split).

\subsection{Augmented Inform Stage}
\label{sec:mem_info_tuple}

During peer network building, each rank sends its local information to $f$ (the
\emph{fanout}) randomly selected peer ranks over a number of asynchronous
\emph{rounds}. When a message is received, its recipient augments it
with its known information and, if the information round is less than
the prescribed number of rounds, propagates it further to $f$ ranks
not visited before by this message, as shown on
line~\ref{alg:line:round_compare} of Figure~\ref{alg:ccm_lb}.

In contrast to previous work~\cite{menon:13}, information beyond rank
loads must be propagated, for without knowledge of the memory and
communication on a given rank, another rank cannot evaluate whether
tasks can be transferred even if it is underloaded.
We thus augment the inform messages with on- and off-rank communication
volumes ($\volon{p}{r}$, $\voloff{p}{r}$), homing cost
$\homemem{p}{r}$, baseline rank footprint memory $\basemem{p}{r}$,
and a summary of the clusters found on that rank~$r$.
For each  cluster $c$, we send the load $\load{p}{c}$, the sizes of
the shared blocks accessed by the cluster $\sharset{p}{c}$, the inter- and
intra-cluster communication volumes ($\volon{p}{c}$, $\voloff{p}{c}$), and the
cluster memory baseline footprint $\basemem{p}{c}$.
This additional data allows us to approximate the work model as we
consider remapping tasks.
We note that while the rank-based additions have constant size, the
cluster-based component is $\mathcal{O}(|C|)$, where $C$ is the set of clusters,
leading to an increased upper-bound on space during the inform stage.

\subsection{Task Transfer Algorithm}
\label{sec:new_algo}

In previous work~\cite{menon:13}, the transfer phase begins after peer network building by
using a cumulative mass function to bias random selection of ranks for potential
transfer depending on how underloaded they are. Then, work units are selected
from the overloaded rank and given to the underloaded without any
intervention. In further work, an underloaded rank was allowed to negatively
acknowledge a requested transfer if it increases the load of this
rank beyond the arithmetic mean of loads across all ranks.

In the landscape of the more complex criterion that includes (1) computation,
(2) communication, and (3) memory as factors, it is infeasible for a rank to
decide unilaterally to transfer tasks without full knowledge of the other rank's
tasks (including full communication information and memory requirements). Thus,
the proposed algorithm operates in two stages. First, it applies the \model
update {formul\ae} to decide how valuable transferring with a target rank might be
(based on information that might be out-of-date). Second, it picks the most
potentially viable rank to transfer work with and attempts to obtain a lock on
that rank. Figure~\ref{alg:ccm_lb} provides an overview of the \dlb
algorithm.

On line~\ref{alg:build_neighborhood} of Figure~\ref{alg:ccm_lb}, we build
the random peer network for each rank, resulting in a list of ranks that each rank
knows about. On line~\ref{alg:apply_criterion}, each rank applies the
criterion to each peer rank by calculating the amount of
work (using the update formul\ae) under possible task transfers. The best
transfer (lowest resulting work) is put in a sorted list (\code{work\_list}) for each
peer. Each rank then tries to lock the peer that has the best potential
transfer (line~\ref{alg:try_lock}).

If all ranks are allowed to request and obtain a lock concurrently, deadlocks
can easily occur. In the simplest case, ranks $r_1$ and $r_2$ might both send
messages requesting a lock from each other. Rank $r_1$ may receive the request
from $r_2$ and allow it to obtain a lock. Concurrently, rank $r_2$ may apply the
same logic and allow $r_1$ to obtain a lock. By the time both ranks are notified,
they may be locked and also hold a lock on the other rank. While a rank is locked
it cannot make progress on a lock it holds because another rank might change its
task distribution. These types of cycles can occur with an arbitrary number.
To remediate this problem, if a rank $r_1$ is locked by
another rank, $r_x$, and also obtains a lock on $r_2$, it immediately releases the lock if
$r_x \leq r_2$ (shown on line~\ref{alg:lock_cycle_avoid}).
This logic guarantees that cycles will not form. The option to lock
$r_2$ is added back to the \code{work\_list} so it can try to obtain the lock later once it
has been unlocked.

Once a lock succeeds, a message is sent from $r_2$ to $r_1$ with up-to-date
information on the tasks residing on $r_2$.
Once received, \code{TryTransfer} is invoked, calling \code{FindBestCCM} (line ~\ref{alg:call_best_ccm}) to search for
clusters that can be given or swapped to reduce the
maximum work between the two ranks.
It evaluates many such possible transfers and selects the best one to execute (if
one exists that is better than the current configuration shown on
line~\ref{alg:config_is_better}).
After every rank has exhausted its list of possible ranks to transfer in \code{work\_list},
and all selected transfers have occurred, the iteration is over. The algorithm proceeds
with the next iteration by creating a new random peer network and performing the
whole process again.

\section{Mixed-Integer Linear Programming Formulation}
\label{sec:milp}
We now reformulate the task-rank problem assignment as a mixed-integer
linear program (MILP), which is NP-hard.
Our \model model proposed in~\S\ref{sec:work_model} makes this formulation
much more complex than a conventional job-machine assignment, due to
the inclusion of additional components with assorted dependencies.
We thus proceed in increasing order of complexity, from constrained
compute-only to the full work model of~\eqref{eqn:work_model}.
\subsection{Common Definitions}
We begin with notational conventions which, albeit not
necessary, shall ease the understanding of our approach.
For instance, summation indices are dummy and can be replaced without
altering the meaning of the sum; but we believe it is more convenient
to the reader if a given index letter always refers to the same type
of entity:
\begin{center}
\setlength{\arrayrulewidth}{1pt}
\begin{tabular}{@{}llcl@{}}\hline
entity type    &set             &cardinality &indices  \\\hline
ranks          &$R$             &$I$         &$i,j$    \\
tasks          &$\taskset{p}{}$ &$K$         &$k,\ell$ \\
communications &$C^p$           &$M$         &$m$      \\
shared blocks  &$\sharset{p}{}$ &$N$         &$n$      \\\hline
\end{tabular}
\end{center}

Subsequently, we define \emph{assignment matrices} between the above
defined entity types, using the indicator function:
\begin{center}
\setlength{\arrayrulewidth}{1pt}
\begin{tabular}{@{}llr@{$\times$}lr@{$:=$}l@{}}\hline
to type &from type     &\multicolumn{2}{c}{size} &\multicolumn{2}{c}{matrix entries} \\[2pt]\hline
tasks   &shared blocks &$K$ &$N$ &$\sbt^p_{k,n}$  &$\mathds{1}_\sharset{p}{t_k}(s_n)$ \\
ranks   &shared blocks &$I$ &$N$ &$\sbh^p_{i,n}$  &$\mathds{1}_\homeset{p}{r_i}(s_n)$ \\
ranks   &shared blocks &$I$ &$N$ &$\phi^p_{i,n}$  &$\mathds{1}_\sharset{p}{r_i}(s_n)$ \\
ranks   &tasks         &$I$ &$K$ &$\chi^p_{i,k}$  &$\mathds{1}_\taskset{p}{r_i}(t_k)$ \\[2pt]\hline
\end{tabular}
\end{center}
Although $(\sbh^p)$ and $(\phi^p)$ have the same
shape, they differ in that the former specifies homing of a
block to a rank, while the latter indicates the presence of the block on a
rank.
The problem statement does not allow
the load balancer to modify either block-task or block-home assignments;
as a result, both $(\sbt^p)$ and $(\sbh^p)$ are parameters,
whereas both $(\phi^p)$ and $(\chi^p)$ are variables, whence the use of
different alphabets to emphasize this contrast.
Meanwhile, every task must be assigned to exactly one rank, and at
most one shared block, while every block shall be homed at exactly one
rank; therefore, the following~$2K+N$ \emph{consistency}
constraints must hold:
\begin{align}
\label{eqn:consistency}
\forallint{k}{K}\enskip
&\sum_{n=1}^{N}\sbt^p_{k,n}\leq1\enskip\land\enskip
\sum_{i=1}^{I}\chi^p_{i,k}=1,\\
\forallint{n}{N}\enskip
&\sum_{i=1}^{I}\sbh^p_{i,n}=1. 
\end{align}

\begin{figure}[htbp]
\begin{mdframed}
\hspace{0.01\columnwidth}
\begin{minipage}[t]{0.62\columnwidth}\vspace{0pt}
\centering
\begin{tabularx}{\textwidth}{@{}Y@{}Y@{}}
$\scriptstyle\taskset{p}{r_1}=\{t_1,t_2\}$
&$\scriptstyle\taskset{p}{r_2}=\{t_3\}$
\end{tabularx}\\[1pt]
\includegraphics[width=0.99\textwidth]{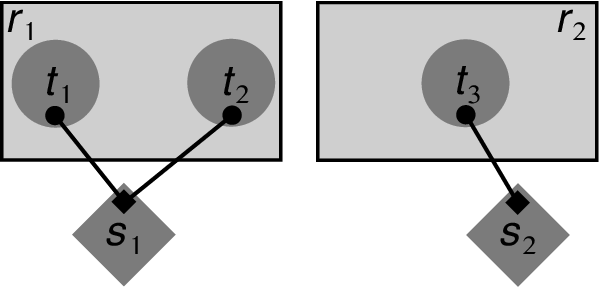}
\end{minipage}
\hfil
\begin{minipage}[t]{0.35\columnwidth}\vspace{2pt}
$\scriptstyle\Rightarrow\enskip
\scriptstyle(\chi^p)=\left(\begin{smallmatrix}1&1&0\\0&0&1\end{smallmatrix}\right)$\\[4pt]
$\begin{array}{@{}s@{}s@{}}
\sharset{p}{t_1} = \{s_1\} &\rvb{3}{-4}{20}{\!\scriptstyle(\sbt^p)=\left(\begin{smallmatrix}1&0\\1&0\\0&1\end{smallmatrix}\right)}\\
\sharset{p}{t_2} = \{s_1\}\\
\sharset{p}{t_3} = \{s_2\}
\end{array}$\\[8pt]
$\begin{array}{@{}s@{\,}s@{\,}l@{}}
(\phi^p) &=
&\scriptstyle\left(\begin{smallmatrix}1&1&0\\0&0&1\end{smallmatrix}\right)
\odot\left(\begin{smallmatrix}1&0\\1&0\\0&1\end{smallmatrix}\right)\\
&= &\left(\begin{smallmatrix}1&0\\0&1\end{smallmatrix}\right)\hfill\\
\end{array}$
\end{minipage}
\caption{\label{fig:lomcp_example}A Compute-Only Memory-Constrained Problem (COMCP) example
for $I$$=$$2$, $K$$=$$3$, and $N$$=$$2$, with corresponding assignment sets and matrices.}
\end{mdframed}
\end{figure}
As an illustrative example, consider an arrangement with
$I$$=$$2$, $K$$=$$3$, and $N$$=$$2$, where the two first tasks are
assigned to the first rank and share a memory block, while the
remaining task is assigned to the second rank and is a lone
participant in a second memory block, as shown in
Figure~\ref{fig:lomcp_example}.
From this we obtain the assignment matrices, with both $(\sbt^p)$ and
$(\chi^p)$ satisfying the consistency constraints
in~\eqref{eqn:consistency}.
We further observe that, although $2^6$$=$$64$ different combinations
of the $6$ binary entries in $(\chi^p)$ may be formed, the
consistency constraint eliminates $3$ degrees of freedom because the
knowledge of, e.g., the first row unambiguously determines the second.
There are, as expected, only $2^3$$=$$I^K$$=8$ consistent
task-rank assignments.
  
Obviously, shared blocks do not roam freely between ranks; rather,
their assignments are constrained by their associated tasks.
Denoting $\odot$ the Boolean product between binary
matrices, i.e., using the arithmetic of the $(\binset,\lor,\land)$
semiring, the fundamental constraint is:
\begin{restatable}[Boolean shared block matrix relations]{theorem}{thmboolblockmat}
\label{pro:boolean_block_rank_task}
\hfill\\[-8pt]
\begin{equation}
\label{eqn:boolean_block_rank_task}
(\phi^p)=(\chi^p)\odot(\sbt^p).
\end{equation}
\end{restatable}
\begin{proof}
Omitted for brevity, refer to Appendix~\ref{sec:proofs}.
\end{proof}
Unfortunately, albeit elegant and tight, this property based on
Boolean algebra does not lend itself to a linear program formulation.
It must instead be re-cast in an integral, yet much less concise form, as
follows:
\begin{restatable}[Integer shared block matrix relations]{theorem}{thmintblockmat}
\label{pro:integer_block_rank_task}
\hfill\\[-8pt]
\begin{empheq}[%
    left=\hspace{-5pt}(\forall(i{,}n)\in\ii{I}\!\times\!\ii{N})\!\empheqlbrace]{align}
\label{eqn:nu_ge_nu_xi}
&\!(\forall k\!\in\!\ii{K})\,\phi^p_{i,n}\geq\sbt^p_{k,n}\chi^p_{i,k}\hspace{-3pt}\\
\label{eqn:nu_le_sum_nu_xi}
&\phi^p_{i,n}\leq\sum_{k=1}^K\sbt^p_{k,n}\chi^p_{i,k}.
\end{empheq}
\end{restatable}
\begin{proof}
Omitted for brevity, refer to Appendix~\ref{sec:proofs}.
\end{proof}
\begin{table}[htb!]
\begin{center}
\setlength{\arrayrulewidth}{1pt}
\newcommand{\lbl}[1]{{{$\hfill{#1}\hspace{14pt}<$}}}
\newcommand{\lbe}[1]{{{$\hfill{#1}\hspace{14pt}=$}}}
\newcommand{\ubl}[1]{{{$<\,\quad{#1}\hfill$}}}
\newcommand{\ube}[1]{{{$=\,\quad{#1}\hfill$}}}
\newcommand{\lsv}[1]{{\multirow{6}{*}[1pt]{${#1}\left\{\rule{0pt}{30pt}\right.$}}}
\newcommand{\ltv}[1]{{\multirow{3}{*}[1pt]{${#1}\left\{\rule{0pt}{15pt}\right.$}}}
\newcommand{\rtv}[1]{{\multirow{3}{*}[1pt]{\hspace{-10pt}$\left.\rule{0pt}{15pt}\right\}{#1}$}}}
\begin{tabular}{@{}l@{\enskip}l@{\enskip}c@{\quad}c@{\quad}c@{\enskip}c@{ }c@{ }c@{}}\hline
$~i$ &$n$ &$k$ &$\sbt^p_{k,n}$ &$\chi^p_{i,k}$
  & $\sbt^p_{k,n}\chi^p_{i,k}\leq$
  & $\phi^{p\vphantom{A}}_{i,n}$
  & $\leq\sum_k\sbt^p_{k,n}\chi^p_{i,k}$                     \\\hline
\lsv{1} &\ltv{1} &$1$ &$1$ &$1$ &\lbe{1} &\rtv{1} &        \\
        &        &$2$ &$1$ &$1$ &\lbe{1} &        &\ubl{2} \\
        &        &$3$ &$0$ &$0$ &\lbl{0} &        &        \\\cline{2-8}\\[-8pt]
        &\ltv{2} &$1$ &$0$ &$1$ &\lbe{0} &\rtv{0} &        \\
        &        &$2$ &$0$ &$1$ &\lbe{0} &        &\ube{0} \\
        &        &$3$ &$1$ &$0$ &\lbe{0} &        &        \\\hline
\lsv{2} &\ltv{1} &$1$ &$1$ &$0$ &\lbe{0} &\rtv{0} &        \\
        &        &$2$ &$1$ &$0$ &\lbe{0} &        &\ube{0} \\
        &        &$3$ &$0$ &$1$ &\lbe{0} &        &        \\\cline{2-8}\\[-8pt]
        &\ltv{2} &$1$ &$0$ &$0$ &\lbl{0} &\rtv{1} &        \\
        &        &$2$ &$0$ &$0$ &\lbl{0} &        &\ube{1} \\
        &        &$3$ &$1$ &$1$ &\lbe{1} &        &        \\\hline
\end{tabular}
\end{center}
\caption{Compliance of assignment matrices to
  constraints~\eqref{eqn:nu_ge_nu_xi} \& \eqref{eqn:nu_le_sum_nu_xi}.}
\label{tab:assignment_constraints}%
\end{table}
We further remark that~\eqref{eqn:nu_ge_nu_xi} provides tight bounds
for all  $K$ inequalities when $\phi_{i,n}$$=$$0$, and at least once when
$\phi_{i,n}$$=$$1$.
The latter can be established by contradiction: assuming strictness of
all inequalities would imply that all $\sbt^p_{k,n}\chi^p_{i,k}$ be nil,
hereby causing $\phi_{i,n}$ to vanish as well and thus contradicting
the hypothesis on $\phi_{i,n}$$=$$1$. 
In contrast,~\eqref{eqn:nu_le_sum_nu_xi} does not provide a tight
upper bound in general, as exhibited by
Table~\ref{tab:assignment_constraints}, which expounds the earlier
illustrative example: for instance, $\phi_{1,1}$$=$$1$
but~\eqref{eqn:nu_le_sum_nu_xi} only provides a loose upper bound
thereof.
However, because this may only occur when $\phi_{i,n}$$=$$1$,
the problem is automatically remedied by the fact that the search
space for $\phi_{i,n}$ is limited to~$\binset$; in other words, another constraint (that of the definition domain) will compensate for
the loose bound.
\subsection{Compute-Only Memory-Constrained Problem (COMCP)}
The aim of this simplified model is to ensure that our approach works
for compute-only load balancing under memory constraint, i.e., with
$\alpha$$=$$1$ and $\beta$$=$$\gamma$$=$$\delta$$=$$0$
in~\eqref{eqn:work_model}.
Regarding~$\varepsilon$, the per-rank bound of~\eqref{eqn:memory_ub}
is used in combination with the definitions of $(\chi^p)$ and
$(\phi^p)$ which, after replacing the $\max$ operator
in~\eqref{eqn:taskmem} with $K$ linear inequalities, equivalently
amounts to the following $I$$\times$$K$ constraints:
\begin{multline}
\label{eqn:membound_base_only}
\!\!\!(\forall(i{,}k)\in\ii{I}\!\times\!\ii{K})\enskip
\sum_{\ell=1}^{K}\basemem{p}{t_\ell}\chi_{i,\ell} + \overmem{p}{t_k}\chi_{i,k}\\
+ \sum_{n=1}^{N}\normmem{p}{s_n}\phi_{i,n}
\leq \avaimem{r_i}\!-\!\basemem{p}{r_i}.
\end{multline}%

It might be tempting to view the decision variable as that obtained
by vectorizing\footnote{\emph{How} this vectorization is performed,
i.e., in row or column-major order, is an implementation detail.} the
$(\chi^p)$ and $(\phi^p)$ assignment matrices, followed by the concatenation
of an $I(K$$+$$N)$-dimensional binary vector.
However, this approach cannot be formulated in terms of a linear
program because its objective function, $\max_{r\in{}R}\workmod{p}{r}$,
is not a linear combination of these binary variables.
This difficulty can be resolved with the \emph{makespan}
formulation~\cite{wagner1959integer}, which introduces an additional
degree of freedom,
in the form of a nonnegative continuous variable~$\workmax{p}$
constraining $\workmod{p}{r}$ from above.
As work is reduced to the compute term in~\eqref{eqn:work_model}, this
can be equivalently formulated in $I$ new constraints using task-rank
assignments:
\begin{equation}
\label{eqn:load_only_constraint}
\forallint{i}{I}\enskip\sum_{k=1}^{K}\load{p}{t_k}\chi_{i,k}\leq\workmax{p}.
\end{equation}
Forming the vector $\vec{x}^p$$=$$\mathrm{vec}\left(\overrightarrow{(\chi^p)},\overrightarrow{(\phi^p)},\workmax{p}\right)$ 
finally allows us to formulate the COMCP as the following MILP:
\begin{alignat}{2}
\label{eqn:argmin_cx}
\textstyle&\argmin_{\vec{x}^p\in\binset^{I(K\!+\!N)}\times\Rpos} \quad && \vec{c}\cdot\vec{x}^p\\
\label{eqn:st_Ax_1}
&\mathrm{subject~to}   \quad && A\vec{x}^p = \vec{1}_K \\
\label{eqn:st_Bx_b_LOMC}
&\mathrm{subject~to}   \quad && B\vec{x}^p +\vec{b} \geq \vec{0}_{I(K\!+\!1)(N\!+\!1)}
\end{alignat}
where $\vec{c}$ has all nil entries, except for a final, unit
coordinate corresponding to the $\workmax{p}$ entry in $\vec{x}^p$, so
that $\vec{c}\cdot\vec{x}^p=\workmax{p}$. 
Meanwhile, matrices $A$ (determined by the part of~\eqref{eqn:consistency}
concerning~$(\chi^p)$) and $B$, and vector $\vec{b}$ (determined
by~\eqref{eqn:nu_ge_nu_xi}, \eqref{eqn:nu_le_sum_nu_xi},
\eqref{eqn:membound_base_only}, and~\eqref{eqn:load_only_constraint})
only contain input parameter values.

Continuing with the running example, $\vec{x}^p$ is $11$-dimensional
and, e.g., using row-major vectorization and column-vector convention,
$A=\left(\begin{smallmatrix}
  1&0&0&1&0&0&0&0&0&0&0\\
  0&1&0&0&1&0&0&0&0&0&0\\
  0&0&1&0&0&1&0&0&0&0&0
\end{smallmatrix}\right)$.
For brevity, we only provide the main principles presiding to assembly of~$B$, whose size is
$24\times11$: its first $16$ are directly given by the inequalities in
Table~\ref{tab:assignment_constraints} ($12$ and $4$ lower and upper 
bounds, respectively).
Denoting parameters $d_\ell$$=$$-\basemem{p}{t_\ell}$, $e_k$$=$$-\overmem{p}{t_k}$,
$f_n$$=$$-\normmem{p}{s_n}$, and $g_k$$=$$-\load{p}{t_k}$ for
conciseness, the last $6$$+$$2$$=$$8$ rows of $B$ are provided
by~\eqref{eqn:membound_base_only} and~\eqref{eqn:load_only_constraint}
as follows:
\begin{equation*}
\left(\begin{smallmatrix}
  d_1+e_1&d_2&d_3&0&0&0&f_1&f_2&0&0&0\\
  d_1&d_2+e_2&d_3&0&0&0&f_1&f_2&0&0&0\\
  d_1&d_2&d_3+e_3&0&0&0&f_1&f_2&0&0&0\\
  0&0&0&d_1+e_1&d_2&d_3&0&0&f_1&f_2&0\\
  0&0&0&d_1&d_2+e_2&d_3&0&0&f_1&f_2&0\\
  0&0&0&d_1&d_2&d_3+e_3&0&0&f_1&f_2&0\\
  g_1&g_2&g_3&0&0&0&0&0&0&0&1\\
  0&0&0&g_1&g_2&g_3&0&0&0&0&1
\end{smallmatrix}\right).
\end{equation*}
Finally, all entries of $\vec{b}$ are nil, except those two
corresponding to the opposite of the right-hand sides
in~\eqref{eqn:membound_base_only}.
\subsection{Full Work Model Problem (FWMP)}
All terms in~\eqref{eqn:work_model} are now retained.
The framework introduced above for the COMCP is conserved; in particular:
\begin{itemize}
\item
the minimization problem~\eqref{eqn:argmin_cx} remains essentially
identical, but $\vec{x}^p$ is expanded to include the assignments of
communications to ranks, while $\vec{c}$ is expanded by as many nil
entries so that only $\workmax{p}$ will not be canceled;
\item
constraint~\eqref{eqn:st_Ax_1} is kept unchanged, and so are the
contributions to~\eqref{eqn:st_Bx_b_LOMC}
of~\eqref{eqn:nu_ge_nu_xi},~\eqref{eqn:nu_le_sum_nu_xi}, 
and~\eqref{eqn:membound_base_only}.
\end{itemize}
However, the communication volumes and homing costs must be added to
$B$ and~$\vec{b}$.
We thus introduce third-order assignment tensors, keeping our
earlier Latin/Greek alphabetical convention for parameters
vs. variables, respectively:
\begin{center}
\setlength{\arrayrulewidth}{1pt}
\begin{tabular}{@{}l@{\;\;}l@{ }r@{$\times$}c@{$\times$}l@{\;\;}r@{$:=$}l@{}}\hline
to type &from type      &\multicolumn{3}{c}{size}
  &\multicolumn{2}{c}{tensor entries}                           \\[2pt]\hline
tasks   &communications &$K$ &$K$ &$M$ &$\ctt^p_{k,\ell,m}$
  &$\mathds{1}_{\sentset{p}{t_k}\!\cap\recvset{p}{t_\ell}}(c_m)$\\
ranks   &communications &$I$ &$I$ &$M$ &$\psi^p_{i,j,m}$
  &$\mathds{1}_{\sentset{p}{r_i}\!\cap\recvset{p}{r_j}}(c_m)$   \\[3pt]\hline
\end{tabular}
\end{center}
We note one peculiarity of these tensors: as a communication edge has
only two endpoints, they have only one nonzero entry per $m$-slice.
The results of Theorem~\ref{pro:boolean_block_rank_task} are
readily extended to the communication assignment tensors:
\begin{restatable}[Boolean communication tensor relations]{theorem}{thmboolcommtens}
\label{pro:boolean_communication_rank_task}
\hfill\\[-8pt]
\begin{equation}
\label{eqn:boolean_identity_tensors}
\forallint{m}{M}\enskip\slc{\psi}{p}{m}
=(\chi^p)\odot\slc{\ctt}{p}{m}\odot\transp{(\chi^p)}.
\end{equation}
\end{restatable}
\begin{proof}
Omitted for brevity, refer to Appendix~\ref{sec:proofs}.
\end{proof}

\begin{figure}[htbp]
\begin{mdframed}
\begin{minipage}[t]{\columnwidth}
$\,\scriptstyle\sentset{p}{t_1}\!=\{c_1,c_4\},
\hfil\sentset{p}{t_2}\!=\{c_2\},
\hfil\sentset{p}{t_3}\!=\{c_3\},
\hfil\recvset{p}{t_1}\!=\varnothing,
\hfil\recvset{p}{t_2}\!=\{c_3,c_4\},
\hfil\recvset{p}{t_3}\!=\{c_1,c_2\}$
\end{minipage}\\[-7pt]
\hspace{0.01\columnwidth}
\begin{minipage}[t]{0.62\columnwidth}\vspace{0pt}
\centering
\includegraphics[width=0.99\textwidth]{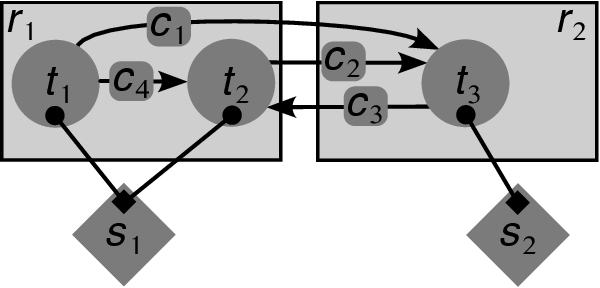}\\[-5pt]
\mbox{$\scriptstyle\sentset{p}{r_1}\!=\{c_1,c_2,c_4\},
\,\sentset{p}{r_2}\!=\{c_3\},
\,\recvset{p}{r_1}\!=\{c_3,c_4\},
\,\recvset{p}{r_2}\!=\{c_1,c_2\}
\,\Rightarrow$}
\end{minipage}
\hfil
\begin{minipage}[t]{0.35\columnwidth}\vspace{0pt}
\centering
$\hspace{-8pt}\scriptstyle\Rightarrow[\ctt^p]$\\[-4pt]
$\overbrace{\hspace{0.9\textwidth}}$\\[-9pt]
\begin{tabularx}{0.9\textwidth}{@{}Y@{}Y@{}Y@{}Y@{}@{}}
\sss{\ctt}{p}{1} &\sss{\ctt}{p}{2} &\sss{\ctt}{p}{3} &\sss{\ctt}{p}{4}
\end{tabularx}\\
$\left[\!
\begin{smallmatrix}0&0&0\\0&0&0\\1&0&0\end{smallmatrix}
\!\middle|\!    
\begin{smallmatrix}0&0&0\\0&0&0\\0&1&0\end{smallmatrix}  
\!\middle|\!    
\begin{smallmatrix}0&0&0\\0&0&1\\0&0&0\end{smallmatrix}  
\!\middle|\!    
\begin{smallmatrix}0&0&0\\1&0&0\\0&0&0\end{smallmatrix}
\!\right]$\\[2pt]
$\scriptstyle(\chi^p)\odot\;\Downarrow\;\odot\transp{(\chi^p)}$
$\left[
\begin{smallmatrix}0&0\\1&0\end{smallmatrix}
\middle|    
\begin{smallmatrix}0&0\\1&0\end{smallmatrix}  
\middle|    
\begin{smallmatrix}0&1\\0&0\end{smallmatrix}  
\middle|    
\begin{smallmatrix}1&0\\0&0\end{smallmatrix}
\right]$\\[-2pt]
\begin{tabularx}{0.7\textwidth}{@{}Y@{}Y@{}Y@{}Y@{}}
\sss{\psi}{p}{1} &\sss{\psi}{p}{2} &\sss{\psi}{p}{3} &\sss{\psi}{p}{4}
\end{tabularx}\\[-11pt]
$\underbrace{\hspace{0.7\textwidth}}$\\
$\scriptstyle[\psi^p]$
\end{minipage}
\caption{\label{fig:fwmp_example}A FWMP example
  for $I$$=$$2$, $K$$=$$3$, $M$$=$$4$, and $N$$=$$2$, with
  corresponding communication assignment sets and tensors.}
\end{mdframed}
\end{figure}%
By adding $M$$=$$4$ inter-task communications to the COMCP example
of~Figure~\ref{fig:lomcp_example},
Theorem~\ref{pro:boolean_communication_rank_task} is illustrated
in Figure~\ref{fig:fwmp_example}.
As done for the assignment matrices, the tensor constraints are
reformulated in integral terms for the MILP framework:
\begin{restatable}[Integer communication tensor relations]{theorem}{thmintcommtens}
\label{pro:boolean_communication_rank_task_to_integer}
\hfill\\[-8pt]
\begin{empheq}[%
    left=\hspace{-6pt}
    \begin{array}{@{}c@{}}
      \forallint{i}{I}\\
      \forallint{j}{I}\\
      \forallint{m}{M}
    \end{array}\!\!\empheqlbrace]{align}
\label{eqn:tau_ub_xi}
\psi_{i,j,m} &\leq \sum_{\ell=1}^L\sum_{k=1}^K\chi_{i,k}\ctt_{k,\ell,m},\\
\label{eqn:tau_ub_xj}
\psi_{i,j,m} &\leq \sum_{\ell=1}^L\sum_{k=1}^K\chi_{j,\ell}\ctt_{k,\ell,m},\\
\label{eqn:tau_lb_xixj}
\psi_{i,j,m} &\geq \sum_{\ell=1}^L\sum_{k=1}^K(\chi_{i,k}\!+\!\chi_{j,\ell})\ctt_{k,\ell,m}\!-\!1.\!
\end{empheq}
\end{restatable}
\begin{proof}%
Omitted for brevity, refer to Appendix~\ref{sec:proofs}.
\end{proof}
We illustrate
Theorem~\ref{pro:boolean_communication_rank_task_to_integer} with
the example of Figure~\ref{fig:fwmp_example};
first,~\eqref{eqn:tau_lb_xixj} immediately sets the values of
$\psi_{i,j,m}$ that are equal to~$1$, 
as they cannot be greater by definition, as shown in bold:
\begin{equation}
\label{eqn:tau_lb_xixj_example}
\left[
\begin{smallmatrix}0&\smo\\\mathbf{1}&0\end{smallmatrix}
\middle|    
\begin{smallmatrix}0&\smo\\\mathbf{1}&0\end{smallmatrix}  
\middle|    
\begin{smallmatrix}0&\mathbf{1}\\\smo&0\end{smallmatrix}  
\middle|    
\begin{smallmatrix}\mathbf{1}&0\\0&\smo\end{smallmatrix}
\right]
\leq
\left[
\begin{smallmatrix}0&0\\\mathbf{1}&0\end{smallmatrix}
\middle|    
\begin{smallmatrix}0&0\\\mathbf{1}&0\end{smallmatrix}  
\middle|    
\begin{smallmatrix}0&\mathbf{1}\\0&0\end{smallmatrix}  
\middle|    
\begin{smallmatrix}\mathbf{1}&0\\0&0\end{smallmatrix}
\right].
\end{equation}
We note that when $\psi^p_{i,j,m}$ must vanish, the lower bounds
in~\eqref{eqn:tau_lb_xixj_example} are unimportant, for the search
space always bounds it from below at~$0$.
In turn,~\eqref{eqn:tau_ub_xi} and~\eqref{eqn:tau_ub_xj} thus respectively
yield 
\begin{equation}
\hspace{-8pt}
\left[
\begin{smallmatrix}\bz&\bz\\1&1\end{smallmatrix}
\middle|    
\begin{smallmatrix}\bz&\bz\\1&1\end{smallmatrix}  
\middle|    
\begin{smallmatrix}1&1\\\bz&\bz\end{smallmatrix}  
\middle|    
\begin{smallmatrix}1&1\\\bz&\bz\end{smallmatrix}
\right]
\geq
\left[
\begin{smallmatrix}\bz&\bz\\1&\bz\end{smallmatrix}
\middle|    
\begin{smallmatrix}\bz&\bz\\1&\bz\end{smallmatrix}  
\middle|    
\begin{smallmatrix}\bz&1\\\bz&\bz\end{smallmatrix}  
\middle|    
\begin{smallmatrix}1&\bz\\\bz&\bz\end{smallmatrix}
\right]
\leq
\left[
\begin{smallmatrix}1&\bz\\1&\bz\end{smallmatrix}
\middle|    
\begin{smallmatrix}1&\bz\\1&\bz\end{smallmatrix}  
\middle|    
\begin{smallmatrix}\bz&1\\\bz&1\end{smallmatrix}  
\middle|    
\begin{smallmatrix}1&\bz\\1&\bz\end{smallmatrix}
\right]\hspace{-8pt}
\end{equation}
thereby setting $\psi^p_{i,j,m}$ to $0$ at least once when needed, as
shown in bold as well.
We see that all entries in $[\psi^p]$ are indeed unambiguously
set by Theorem~\ref{pro:boolean_communication_rank_task_to_integer}.

Finally, the continuous constraints~\eqref{eqn:load_only_constraint}
in the COMCP with the following, including all terms
in~\eqref{eqn:work_model}:
\begin{multline}
\label{ilp:full_model_constraint}
(\forall(i{,}\sigma)\in\ii{I}\!\times\!\mathfrak{S}_2)\enskip  
\alpha\sum_{k=1}^{K}\load{p}{t_k}\chi_{i,k}\\
+\beta\sum_{m=1}^{M}\sum_{j=1\atop{j\neq{}i}}^{I}\voloff{p}{c_m}\psi^p_{\sigma(i,j),m}
+\gamma\sum_{m=1}^{M}\volon{p}{c_m}\psi^p_{i,i,m}\\
+\delta\sum_{n=1}^{N}\normmem{p}{s_n}(1-\sbh^p_{i,n})\phi^p_{i,n}
\leq\workmax{p}.
\end{multline}
For the sake of brevity, we make only a few observations:
\begin{itemize}
\item
the compute term is unchanged from~\eqref{eqn:load_only_constraint},
except for its multiplication by~$\alpha$ as required
by~\eqref{eqn:work_model};
\item
the off-node communication ($\beta$) term is derived
from~\eqref{eqn:rank_agg_comm} but, because the $\max$ operator is
non-linear, it is replaced with one upper bound constraint for each
operand; as a result, one inequality must be generated for both
permutations of $\{i,j\}$: one for $\psi_{i,j,m}$ and one for
$\psi_{j,i,m}$; 
\item
the on-node communication ($\gamma$) term, obtained
from~\eqref{eqn:rank_agg_comm}, contains only the diagonal
entries of the~$(\psi^p)$ $m$-slices;
\item
and the homing ($\delta$) term results from the fact that
\begin{equation*}
\mathds{1}_{\sharset{p}{r}\setminus\homeset{p}{r}}
=\mathds{1}_{\sharset{p}{r}\cap\homeset{p\complement}{r}}
=\mathds{1}_{\sharset{p}{r}}\times\mathds{1}_{\homeset{p\complement}{r}}
=\mathds{1}_{\sharset{p}{r}}\times(1-\mathds{1}_{\homeset{p}{r}}).
\end{equation*}
\end{itemize}
The above results in $I[(K$$+$$1)(N$$+$$1)$$+$$3IM$$+$$1]$ inequality constraints
for the FWMP, i.e., when using the example of Figure~\ref{fig:fwmp_example}:
$2[(3$$+$$1)(2$$+$$1)$$+$$3$$\times$$2$$\times$$4$$+$$1]$$=$$74$.

%% file: ccm-lb-segment-1.tex
\begin{lstlisting}
info_known = dict() /* rank-local peer network info */

def ComputeCCM(rank, add_tasks = [], remove_tasks = []):
    /* apply update formulae to compute new work with add_tasks added and remove_tasks removed from rank */

def FindBestCCM(rank, peer):
  best_work_diff = -inf
  work_r, work_p = ComputeCCM(rank), ComputeCCM(peer)
  max_work = max(work_r, work_p)
  for c_r in getClusters(rank):
    for c_p in getClusters(peer):
      work_r_after = ComputeCCM(rank, c_p, c_r)
      work_p_after = ComputeCCM(peer, c_r, c_p)
      max_work_after = max(work_r_after, work_p_after)
      work_diff = max_work - max_work_after
      if work_diff > 0:|\label{alg:config_is_better}|
        best_work_diff = max(best_work_diff, work_diff)
  return best_work_diff

def TryTransfer(rank, peer):
  best_work_diff = FindBestCCM(rank, peer)|\label{alg:call_best_ccm}|
  if best_work_diff > 0:
    /* perform task transfers for best_work_diff */

def BuildPeerNetwork(k_rounds, fanout):
  info_known.clear()
  info_known[rank] = /*information from this rank*/
  def spreadInfo(cur_round, new_info):
    info_known[a] = b for a,b in new_info
    if cur_round < k_rounds:|\label{alg:line:round_compare}|
      for f in fanout:
        p = /* generate random peer */
        send(spreadInfo, @p, cur_round+1, info_known)
  spreadInfo(1, nil)
  return info_known.keys()

def CCM_LB(n_iter, k_rounds, fanout, rank):
  for i in n_iter:
    peers = BuildPeerNetwork(k_rounds, fanout)|\label{alg:build_neighborhood}|
    for p in peers:
      work_list.append(FindBestCCM(rank,p),p)|\label{alg:apply_criterion}|
    for (work, peer) in sort(work_list):
      has_lock = TryLock(peer)|\label{alg:try_lock}|
      if has_lock:
        if IsLocked(rank) and GetLockingRank(rank) <= peer:|\label{alg:lock_cycle_avoid}|
          Unlock(peer)
          work_list.append((work, peer))
        else:
          while IsLocked(rank): pass /* wait to be unlocked */
          @when recvUpdate(peer_info):
              info_known[peer] = peer_info /* update */
              TryTransfer(rank, peer)
              Unlock(peer)
\end{lstlisting}

%% file: 50.tex
\section{Application}
\label{sec:application}
\subsection{The \gemma Code}
\label{sec:gemma}
\gemma is developed in support of modernization
activities~\cite{modernization}, to gain insight into how the energy
from electromagnetic interference couples into systems and what
effects can occur as a result.
It uses frequency-domain analysis to solve general EM scattering and
coupling problems, with particular emphasis on performance across a
variety of computing platforms as well as allowing efficient incorporation
of specialized models relevant to the physics under consideration.

\gemma uses the method of moments \cite{harrington_1989} to solve surface
integral equations imposed on the surfaces of the problem structure.
Surfaces are discretized with triangular meshes, while equations and
solution space are discretized using the Rao-Wilton-Glisson basis functions
\cite{rao_wilton_glisson_1982} defined on the mesh triangles.
Testing of the integral equations is performed using an inner product
with these same basis functions (i.e., a Galerkin testing scheme is
used).
The core of the numerical analysis is the construction and solution of
a matrix equation, where each matrix entry is constructed by using the
basis associated with one degree of freedom to test the field radiated
by the basis associated with some (possibly other) degree of freedom.
Any two degrees of freedom associated with elements touching the same
problem region (e.g., triangles that are both on the inner wall of a
cavity structure) will generally be associated with nonzero matrix
entries.

Computing the matrix entries is complicated by the singular nature of
the Green’s function.
Thus, matrix entries involving interactions between nearby
degrees of freedom are more computationally expensive than
others.
When the matrix is partitioned into blocks for parallel computation on
multiple ranks, this discrepancy results in a significant load
imbalance among the different ranks. In coupling problems, this
imbalance is increased by the presence of the zero blocks that occur
due to degrees of freedom not radiating in shared regions.

For our experiments, we run a scaled-up version of the
\texttt{yaml\_rect\_cavity\_2\_slots\_curve} problem from the regression
test suite in \gemma.
This problem mimics the topology of coupling problems of interest 
(e.g., the Higgins cylinder).
The geometry is a 1.8m cubic cavity inside a 2m block of perfectly
conducting metal.
The inner region is connected with the outer by two slots 30cm long,
modeled with aperture width of 0.508mm and a 
depth of 6.35mm. The exterior surface is excited by a plane wave. The inner and
outer slot apertures are discretized with a string of bar elements along each.
We scale up the size of the problem by decreasing the mesh edge length.

\subsection{Overdecomposing into Shared Blocks and Tasks}
\label{sec:overdecomp}
The solver that runs after matrix assembly prescribes how the matrix
must be block-decomposed across MPI ranks.
To allow load balancing of the matrix assembly, we must overdecompose
the assembly work on each MPI rank and make it possible for other
ranks to contribute toward performing that work.
We start by breaking the matrix block on each MPI rank into
\emph{slabs} of contiguous memory.
Due to the layout of the matrix, a slab contains all rows assigned to
the rank but only a subset of the columns.
In the context of our CCM model, each slab corresponds to a shared
memory block; by default, the tasks accessing that slab are co-located
on its home rank.

Each row or column in the matrix corresponds to an unknown.
We overdecompose the work to assemble each shared memory block by
limiting the number of unknowns that will be assigned to a task.
With a limit of $u$, then, a task would contribute to at most a $u$
row by $u$ column subset of the shared block.
When there is more than one type of element in a shared block,
separate tasks are used to compute the contributions from different
element type pairs.
The number of interactions that apply to the unknowns assigned to a
task is pre-computed before the task is instantiated.
In the case where there is no coupling in the unknowns represented
(i.e., a task would not produce any non-zero values), the task is
never instantiated.

\subsection{Finishing the Assembly Process}
\label{sec:finalassembly}
All tasks executed on a given MPI rank that are assigned to the same shared
block will contribute to the same contiguous chunk of memory. Allocating that
chunk of memory is deferred until right before the first task begins
contributing to it, provided that such a task exists. After all matrix
assembly tasks have completed, any shared blocks computed in whole or in part
away from the home rank must be transferred home.

During this transfer process, we must continue to respect memory constraints.
We do so by transferring shared blocks home in waves, limiting at all times the
shared blocks that can exist across each compute node rather than each rank.
Our transfer schedule is not optimal but minimizes situations where two ranks
need to swap shared blocks but lack the memory with which to do so. When such
a situation does occur, one shared block is temporarily moved to a different
compute node with available memory in order to free up memory for the other
transfer to occur. For \gemma, because the cost of the transfers is so much smaller than
the time to compute the shared blocks, we have not attempted to optimize
our algorithm.

Once all instantiated shared blocks are on their home ranks, the memory pages
from the separate shared blocks can be moved into a larger allocation that
represents the matrix block expected by the solver. If a shared block
has not yet been instantiated due to containing all zeros, those zeros are
finally allocated during this process. Deferring those allocations freed up
memory for balancing the workload earlier in the process.

\subsection{Predicting Task Compute Times}
\label{sec:modeling}
In order to load balance \gemma with our approach, the compute times
for the matrix assembly tasks must be predicted.
As \gemma is often run near memory limits, careful orchestration of
the task placement while considering shared matrix blocks is required
to keep the application under memory limits.
Since a persistence-based model from which to derive timing
predictions is not applicable to \gemma, we propose using 
an artificial neural network~\cite{Hornik1989} to approximate the
task-time mapping function.
To build a general model, we ran \gemma on a diverse set of
configurations with a range of interacting element types.
We collect inputs for each task, such as the types of elements, and
fed these into the model.
\subsubsection{Data Pre-Processing Strategy}
\label{sec:data_preprocessing}
When collecting task data across problem configurations, we noticed there are
many more short-duration tasks than longer duration ones. Thus, we developed a
dynamic data point reduction algorithm (detailed in
Appendix~\ref{appendix:data_reduction}) that utilizes a bin decomposition
approach to randomly eliminate points from over-represented data segments.
Additionally, we employed a standard scaler to normalize the training data,
ensuring that each feature contributes equally to the model by having a mean of
zero and unit variance, which is crucial for the effective training and
convergence of the neural network.
\subsubsection{Model Architecture \& Training}
To build the model for the task-time prediction, we employ a
feed-forward neural network (FNN) with 4 hidden layers, each comprised of 200
neurons.  To maintain input distribution consistency across layers, we use batch
normalization on the hidden layers, ensuring stable
training~\cite{ioffe2015batchnorm}.  In order to avoid overfitting the model,
notably towards over-represented smaller tasks, we
utilize~\emph{dropout}~\cite{srivastava2014dropout}---randomly deactivating
neurons during training.
We used the \textit{Leaky} Rectified Linear Unit (ReLU)~\cite{xu2015empirical} for our neuron activation function:
\begin{equation}
  f(x) = x\times\mathds{1}_\Rpos(x)
  + 0.01x\times\mathds{1}_\Rneg(x).
\end{equation}
To train this neural network, we perform the objective function optimization
using \emph{AdamW}, an improved version of the Adam optimizer, as it has been
shown to lead to better generalization and
convergence~\cite{loshchilov2019adamw}.  We utilize the mini-batch method, which
processes data subsets at each iteration of this optimization algorithm to
balance computational efficiency with smoother training.
\subsubsection{Loss Functions}
The standard approach to measuring prediction errors is to compute
either the root mean-squared error (RMSE) or the mean absolute error
(MAE) between the $d$-dimension vector $\vec{p}$ of ground truth
values, and the corresponding vector $\vec{g}$ of predictions.
However, because load imbalance is likely to be more adversely impacted
by over-predicted than under-predicted compute times, we devised an
\emph{under-penalized RMSE}, depending on a nonnegative
parameter $\alpha$, and defined as
$\sqrt{\sfrac{\norm{2}{\err(\vec{p},\vec{g})}^2}{d}}$, where
$\err(\vec{p},\vec{g})$ is the vector of under-penalized errors:
\begin{equation}
\label{eq:upen_errors}
\!\!\forallint{i}{d}\enskip{\err}_i(\vec{p},\vec{g}) =
\begin{cases}
(\vec{g}_i - \vec{p}_i)^2 & \text{if}\;\vec{g}_i - \vec{p}_i \ge 0,\\
\alpha(\vec{g}_i - \vec{p}_i)^2 &\text{otherwise}.
\end{cases}
\end{equation}
In response to this, the trained model barely over-predicts, at the
price of additional under-prediction and, at the time of writing,
empirically produces the best results.

%% file: 70.tex
\section{Empirical Results}
\label{sec:empirical_results}

\begin{figure}
\centering
\setlength{\arrayrulewidth}{1pt}
\subfloat[For a single Gurobi solve at each $\delta$, we show the gap obtained and
the solve time needed or allowed (see text). For twelve \dlb solves at each
$\delta$, we show the min/max of the gap and percent increase in
$\workmax{}$ from the Gurobi solution. The mean \dlb solve time was below
0.7~seconds for all $\delta$.
\label{fig:ccm_ilp_timings}]{%
\footnotesize
\begin{tabular}{@{}l@{\qquad}c@{\quad}c@{\qquad}c@{\quad}c@{}}
& \multicolumn{2}{c}{
  \hspace{-5mm}$\overbrace{\hspace{20mm}}^{\text{\footnotesize\bf Gurobi (MILP)}}$}
& \multicolumn{2}{l}{
  $\overbrace{\hspace{32mm}}^{\text{\footnotesize\bf\dlb}}$}\\[-2pt]
\textbf{Model}    & \textbf{Gap} & \textbf{Solve} & \textbf{Gap}      & \textbf{$\workmax{}\uparrow$}  \\
\phantom{mm}\textbf{$\delta$} &              & \textbf{Time}  & \textbf{Min / Max}& \textbf{Min / Max} \\\hline
1e-9              & 1.2e-3       & 46h 41m        & 6.7e-4 / 1.1e-2 & -0.1\% / 1.0\%         \\
1e-10             & 2.4e-4       & 8h             & 9.9e-3 / 1.8e-2 & 1.0\% / 1.8\%          \\
1e-11             & 1.0e-4       & 8h             & 8.1e-3 / 1.9e-2 & 0.8\% / 1.8\%          \\
1e-12             & 8.0e-5       & 29s            &    --            & --                    \\
1e-13             & 7.9e-5       & 128s           &    --            & --                    \\
0                 & 9.8e-5       & 498s           & 9.0e-3 / 1.8e-2  &0.9\% / 1.8\%          \\
\hline
\end{tabular}%
}\\
\subfloat[Blocks that need to be sent home, the time required for such
transfers, and the compute time for each $\delta$. \gemma was run twelve times
on the single Gurobi solution, and once on each of the twelve \dlb solutions.
\label{fig:transfer_unhomed}]{
  \includegraphics[width=0.45\textwidth]{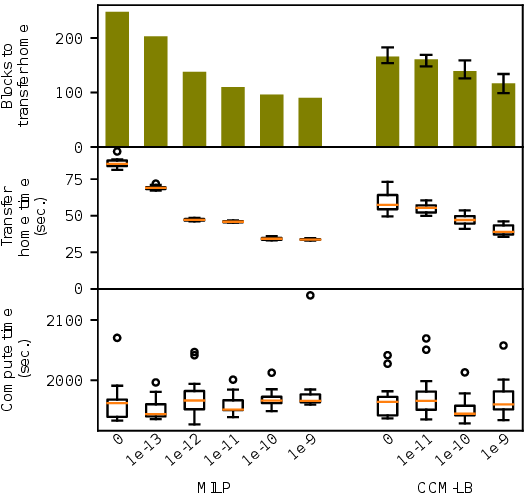}
  \label{fig:ilp_ccmlb_graphs}
}
   \caption{Results comparing the Gurobi (MILP) solutions to \dlb.}
\vspace{-8pt}
\end{figure}

All experiments were performed on a cluster with 1488 nodes connected with Intel
Omni-Path, each composed of 2.9 GHz Intel Cascade Lake 8268 processors with 192
GiB RAM/node. We used two ranks per node (one per socket), with 24 threads each
bound to a core on the socket.

\subsection{MILP and \dlb Comparison}
\label{sec:ilp_comparison}
We have implemented the FWMP (cf.~\S\ref{sec:milp}) for a small-scale \gemma
problem using the PuLP~\cite{mitchell2011pulp} library in Python, which outputs a
general LP format that can be run with different solvers. From initial testing,
we found that the commercial solver, Gurobi~\cite{gurobi}, far exceeded
open-source solvers (such as CBC~\cite{forrest2005cbc} or
GLPK~\cite{makhorin2008glpk}) in speed and quality of solution, and so it was used
as the comparison for our distributed load balancer.

This \gemma experimental problem contains 238,738 unknowns across
14 ranks, 1959 tasks, and 206 shared memory blocks with non-zeros.
For this set of small-scale experiments, we used the
actual task timings from a prior \gemma run, rather than our model
described in \S\ref{sec:modeling}, to reduce the impact of prediction
inaccuracies when comparing the two approaches.
We note that significant machine noise still results in the task timings
deviating from the previous timings.

The Gurobi solver first solves the LP relaxation problem (feasible in
polynomial time), whereby the integral constraints are relaxed to 
real numbers between $[0,1]$.
This relaxed, continuous solution is a lower bound on the best
possible integral one.
Thus, the \emph{gap} is defined as the relative error between the
minimized amount of work $\workmax{i}$ and the continuous lower bound
$\workmax{l}$, i.e.,
$\sfrac{(\workmax{i}-\workmax{l})}{\workmax{l}}$.

Figure~\ref{fig:ccm_ilp_timings} describes the solutions obtained by the MILP
and \dlb for varying input values of~$\delta$.
We configured the solver with a maximum gap of $10^{-4}$, and the MILP
columns show the gap that Gurobi found along with the time taken.
For the largest value of $\delta$, after more than 46 hours, Gurobi
had not found a solution with a gap below $10^{-4}$, suggesting 
it may not be tractable.
We thus decided to terminate this solve, and to time-limit runs to 8
hours for the remaining values of~$\delta$.
Gurobi ran much faster for smaller values of $\delta$, indicating that
the complexity of solving the MILP highly depends on input parameter values.
The \dlb columns show the ranges of the gaps and the relative
increases in $\workmax{}$, compared to the best Gurobi solution found.
\dlb was solved twelve times for each $\delta$, coming up with a
different solution each time, and we stopped decreasing $\delta$ once
it was clear that the transfer time was near that of
$\delta$$=$$0$.
The mean time for \dlb to find a solution was always
under $0.7$~seconds, running online and much faster than the MILP solver.

For all values of $\delta$ except for the largest one, the Gurobi solution
gets closer to the optimal (smaller gap), which is expected as
\dlb is a heuristic-based approach.
Interestingly, the heuristic-based \dlb in one case finds a better
solution that the MILP one ($-0.1\%$) when the FWMP starts to become
intractable for Gurobi ($\delta$$=$$10^{-9}$).

Figure~\ref{fig:ilp_ccmlb_graphs} depicts the number of blocks $n_{\mathrm{off}}$
that are computed off the home rank, the communication time required
to send them home, and the compute time for the same \gemma case, with
varying~$\delta$ in the \model model. 
As $\delta$ increases, $n_{\mathrm{off}}$ should decrease (and the required
communication time as well), and this is indeed what we observe: a
strong inverse correlation between $\delta$ and 
$n_{\mathrm{off}}$, demonstrating the validity and efficacy of our model and
the relevance of the MILP formulation (FWMP).
We also observe that~$\delta$ has a larger impact on the
number of blocks to be homed, and thus on the homing cost, with the
MILP solutions than with \dlb; this is because the latter starts from an
existing, co-localized configuration, whereas the former computes a
solution \emph{ab initio}.
Moreover, we note that compute times across varying $\delta$ and gaps
remain within  machine noise, despite the fact that gaps are on
average slightly worse with \dlb.

\subsection{Larger-scale Experiments}
\label{sec:larger_scale}
\begin{figure}
\centering
\renewcommand{\arraystretch}{1.2}
\setlength{\arrayrulewidth}{1pt}
\subfloat[Description of the weak-scaled problem. Shared blocks only
includes those containing non-zero values.\label{fig:weak_scaling_prob}]{%
\footnotesize
\begin{tabular}{@{}c@{\qquad}c@{\qquad}c@{\qquad}c@{}}
\hline
\textbf{Ranks} & \textbf{Unknowns} & \textbf{Tasks} & \textbf{Shared blocks} \\\hline
16  & 243,079 & 2,383  & 286 \\
64  & 488,381 & 8,955  & 896 \\
256 & 983,881 & 34,709 & 3,076 \\
\hline
\end{tabular}%
}\\
\subfloat[
\textsf{\scriptsize{A}}: baseline \gemma (does not support load balancing);
\textsf{\scriptsize{B}}: overdecomposed \gemma without load balancing;
\textsf{\scriptsize{C}}: overdecomposed \gemma with \dlb and $\delta$$=$$10^{-9}$.
The above-bar multiplicative factors are the full assembly
speedups compared to \textsf{\scriptsize{A}} for the same number of ranks.
\label{fig:assembly_weak_scaling}]{
  \centering
  \includegraphics[width=0.45\textwidth]{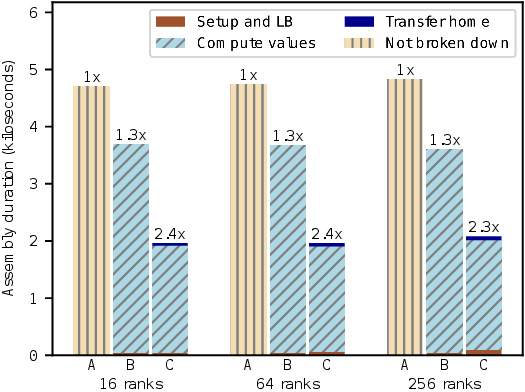}
}
\caption{Speedup of the assembly at each scale.}
\end{figure}

In order to demonstrate our approach at larger scale with a more
realistic case, we weak-scaled a similar \gemma problem to three
different rank counts, as illustrated in
Figure~\ref{fig:weak_scaling_prob}.
For these experiments, we created our neural-network model using
PyTorch, a popular machine learning framework in Python.
We exported trained model weights that can be used in C++ using 
PyTorch's libtorch API~\cite{paszke2019pytorch}.
The model weights were trained using measurements from a version of \gemma
built with an older software stack that exhibited worse computational
performance than the configuration used for these benchmarks.
As a result, \dlb used task time predictions that somewhat differed
from the actual, experimental ones;
we expect better speedups after retraining the
model with the new software stack.

Figure~\ref{fig:assembly_weak_scaling} shows the performance we
attain, with the ``baseline'' corresponding to the unchanged \gemma
code.
By overdecomposing the matrix, we see about a 1.3x speedup due to
cache effects from a different kernel size and zero blocks being
exposed in the matrix (thus less work being performed).
By running our distributed load balancer \dlb on the overdecomposed
code with tasks, we obtain a 2.3-2.4x speedup up to 256 ranks on the
matrix assembly.

%% file: 80.tex
\section{Concluding Remarks}
\label{sec:conclusion}
Using our proposed reduced-order model (\model) for describing
\emph{work} in a parallel system with memory constraints, we have
demonstrated that our distributed load balancer can find
close-to-optimal solutions, leading to substantial speedups in
practice for our target application.
Because we believe that distributed load balancing research
(especially with complex models) should be backed by comparisons to
provably-optimal solutions, we have formulated the optimization
problem as a mixed-integer linear program.
This has allowed us to gauge the optimality of our heuristic-based 
approach and to demonstrate its veracity.
In the future, we plan to show that our model is widely applicable to
many parallel algorithms.

%% file: 100.tex
\section{Proofs of Theorems}
\label{sec:proofs}
\thmhomingupdate*
\begin{proof}
Denoting~$\uplus$ the union of two disjoint sets, and because $A$ and
$B\setminus{A}$ disjoint with union equal to $A\cup{B}$, we have:
\begin{gather}
\!\!\!\sharset{p}{r_1}\!
=\cup_{u\in\taskset{p}{r_1}\!\setminus\!\{t\}}\sharset{p}{u}\cup\sharset{p}{t}
=\newsharset{p}{r_1}\!\cup\sharset{p}{t}
=\newsharset{p}{r_1}\!\uplus(\sharset{p}{t}\!\setminus\!\newsharset{p}{r_1}),\\
\!\newsharset{p}{r_2}\!
=\cup_{u\in\taskset{p}{{r_2}}}\sharset{p}{u}\cup\sharset{p}{t}
=\sharset{p}{r_2}\!\cup\sharset{p}{t}
=\sharset{p}{r_2}\!\uplus(\sharset{p}{t}\!\setminus\!\sharset{p}{r_2}).
\end{gather}
Thus, by right-distributivity of set difference over set union,
\begin{align}
\label{eqn:new_sharset_src}
\sharset{p}{r_1}\!\setminus\!\homeset{p}{r_1}\!
&=(\newsharset{p}{r_1}\!\setminus\!\homeset{p}{r_1})\uplus
((\sharset{p}{t}\!\setminus\!\newsharset{p}{r_1})\!\setminus\!\homeset{p}{r_1}),\\
\label{eqn:new_sharset_dst}
\newsharset{p}{r_2}\!\setminus\!\homeset{p}{r_2}\!
&=(\sharset{p}{r_2}\!\setminus\!\homeset{p}{r_2})\uplus
((\sharset{p}{t}\!\setminus\!\sharset{p}{r_2})\!\setminus\!\homeset{p}{r_2}).
\end{align}
Meanwhile, by definition in~\eqref{eqn:homing_model},
\begin{align}
\newhomemem{p}{r_1} &:= \textstyle
\sum_{s\in\newsharset{p}{r_1}\!\setminus\homeset{p}{r_1}}\!\normmem{p}{s},\\
\newhomemem{p}{r_2} &:= \textstyle
\sum_{s\in\newsharset{p}{r_2}\!\setminus\homeset{p}{r_2}}\!\normmem{p}{s},
\end{align}
whence the conclusion, thanks to the disjoint set unions
in~\eqref{eqn:new_sharset_src} and~\eqref{eqn:new_sharset_dst} that
entails additivity of their memory contents.
\end{proof}
\thmboolblockmat*
\begin{proof}
By definition of
$\sharset{p}{r_i}=\cup_{t_k\in\taskset{p}{r_i}}\sharset{p}{t_k}$
(cf.~\ref{sec:shared_blocs}),
\begin{align}
\label{eqn:block_rank_assignment}
\phi_{i,n}=1
&\iff s_n\in\cup_{t_k\in\taskset{p}{r_i}}\sharset{p}{t_k}\\
&\iff(\exists k\in\ii{K})
\enskip t_k\in\taskset{p}{r_i}\,\land\,s_n\in\sharset{p}{t_k}\\
&\iff\bigvee_{k=1}^{K}\big(\mathds{1}_\taskset{p}{r_i}(t_k)=1\,
\land\,\mathds{1}_\sharset{p}{t_k}(s_n)=1\big)\\
&\iff\bigvee_{k=1}^{K}\big(\chi^p_{i,k}=1\,\land\,\sbt^p_{k,n}=1\big)\\
\label{eqn:exists_k_prod_unit}
&\iff\bigvee_{k=1}^{K}\chi^p_{i,k}\land\sbt^p_{k,n}\,=\,1.
\end{align}
Negating both sides of~\eqref{eqn:exists_k_prod_unit}
yields:%
\begin{equation}
\label{eqn:block_rank_no_natch}
\phi^p_{i,n}=0\iff
\bigvee_{k=1}^{K}\chi^p_{i,k}\land\sbt^p_{k,n}\,=\,0.
\end{equation}
Therefore, as $\phi^p_{i,n}\!\in\binset$, it follows that
\begin{equation}
\label{eqn:boolean_equality_nu_xi_mu}
(\forall(i{,}n)\in\ii{I}\!\times\!\ii{N})\enskip
\phi^p_{i,n}=\bigvee_{k=1}^{K}\chi^p_{i,k}\land\sbt^p_{k,n},  
\end{equation}
whencefrom $(\phi^p)=(\chi^p)\odot(\sbt^p)$ ensues.
\end{proof}
\thmintblockmat*
\begin{proof}
Given any binary values $a$ and $b$, $a\land{}b=ab$, whereas
$a\lor{}b=a+b-ab\leq{}a+b$, which, when applied
to~\eqref{eqn:boolean_equality_nu_xi_mu},
yields~\eqref{eqn:nu_le_sum_nu_xi}.
The proof of~\eqref{eqn:nu_ge_nu_xi} is done by disjunction
elimination:
if $\phi_{i,n}=0$, it comes from~\eqref{eqn:block_rank_no_natch} that 
both sides in~\eqref{eqn:nu_ge_nu_xi} always vanish, making
all $K$ inequalities true.
If $\phi_{i,n}\!=\!1$,~\eqref{eqn:nu_ge_nu_xi} also always holds,
irrespective of the value of its right-hand side, which is
in~$\binset$; all $K$ inequalities are thus true.
\end{proof}
\thmboolcommtens*
\begin{proof}
For brevity, we only provide a notional proof, for it is essentially
similar to that of Theorem~\ref{pro:boolean_block_rank_task}.
Applying the same logic, first to the left Boolean product, transforms
each slice $\slc{\ctt}{p}{m}$, a $K\!\times\!K$ task-to-task matrix,
into an $I\!\times\!K$ task-to-rank matrix, itself by the right
product into an $I\!\times\!I$ rank-to-rank matrix, the
$\slc{\psi}{p}{m}$.
\end{proof}
\thmintcommtens*
\begin{proof}
For all $(i,j)$ and $m$, in $\ii{I}^2$ and $\ii{M}$, respectively,
Theorem~\ref{pro:boolean_communication_rank_task} entails that:
\begin{equation}
\label{eqn:boolean_tensor_entries}
\psi_{i,j,m} =
\bigvee_{\ell=1}^K\bigvee_{k=1}^K\chi_{i,k}\land\ctt_{k,\ell,m}\land\chi_{j,\ell}
\end{equation}
As mentioned above, for a given $m$, $\ctt_{k,\ell,m}$ vanishes for
all $k$ and $\ell$, except for a unique couple $(k_m,\ell_m)$
corresponding to the task endpoints $t_{k_m}$ and $t_{\ell_m}$ of the
directed communication edge $c_m$, for which
$\ctt_{k_m,\ell_m,m}$$=$$1$.
Thus, all but one term in the right-hand side
of~\eqref{eqn:boolean_tensor_entries} vanish, and
\begin{equation}
\psi_{i,j,m}=
\chi_{i,k_m}\land1\land\chi_{j,\ell_m}
= \chi_{i,k_m}\chi_{j,\ell_m}.
\end{equation}
The same logic applies to the double sum over $k$ and $\ell$:
\begin{equation}
\label{eqn:integer_tensor}
\sum_{\ell=1}^K\sum_{k=1}^K\chi_{i,k}\chi_{j,\ell}\ctt_{k,\ell,m}
= \chi_{i,k_m}\chi_{j,\ell_m} = \psi_{i,j,m}.
\end{equation}
Although strictly equivalent in integer terms
to~\eqref{eqn:boolean_identity_tensors}, \eqref{eqn:integer_tensor} is 
not suitable for MILP due the non-linearity in the~$\chi$
components in the decision vector~$\vec{x}^p$ but, by applying the
technique first described in~\cite{glover1974converting}, we
obtain~\eqref{eqn:tau_ub_xi} and~\eqref{eqn:tau_ub_xj} by 
alternatively bounding each of these components from above by~$1$ and,
because
$\mathds{1}_{A\cap{}B}$$+$$\mathds{1}_{A\cup{}B}$$=$$\mathds{1}_A$$+$$\mathds{1}_B$,
\begin{equation}
\chi_{i,k_m}\chi_{j,\ell_m}
\!\!=\chi_{i,k_m}\!\!+\chi_{j,\ell_m}\!\!-\chi_{i,k_m}\!\!\lor\chi_{j,\ell_m}
\!\ge\chi_{i,k_m}\chi_{j,\ell_m}\!\!-1\!
\end{equation}
which, combined with~\eqref{eqn:integer_tensor} and the fact that all
$\ctt_{i,k,m}$, aside from $\ctt_{k_m,\ell_m,m}$$=$$1$,
vanish in the double sum, yields~\eqref{eqn:tau_lb_xixj}.
\end{proof}

%% file: 90.tex
\section{Data Reduction Algorithm}
\label{appendix:data_reduction}
Consider a dataset, arranged in a table $\mathcal{T}$ with $n_r$ rows
of individual observations.
Given a target number of rows $n_r^\ast$, a number of bins $n_b$, and
a \emph{downsampling coefficient} $\theta\!\in]0,1[$, our method is
described in Algorithm~\ref{alg:data_reduction}.
\begin{algorithm}[htbp]
\begin{algorithmic}[1]
\Function{DynamicDataReduce}{$n_r$, $\mathcal{T}$, $n_b$, $\theta$}
  \State $H\gets$\Call{Histogram}{$n_b$,$\mathcal{T}$}\Comment{$n_b$ bins of rows}
  \State $n_r^{\scriptscriptstyle-} \gets n_r - n_r^\ast$
  \Comment{number of bins to drop}
  \While{$n_r^{\scriptscriptstyle-} > 0$}
  \State $b_{\max}\gets\argmax_{b\in{}H}$ \Call{NumberOfRows}{$b$}
  \State $n_{\max}\gets$ \Call{NumberOfRows}{$b_{\max}$}
  \State $n\gets\min([\theta\times{}n_{\max}],n_r^{\scriptscriptstyle-})$
  \State Randomly remove $n$ rows from~$b$.
  \State $n_r^{\scriptscriptstyle-} \gets n_r^{\scriptscriptstyle-} - n$
  \EndWhile
\EndFunction
\end{algorithmic}
\caption{Dynamic data point reduction.}
\label{alg:data_reduction}
\end{algorithm}
The role of~$\theta$ is to arbitrage between convergence speed, and need
to obtain a well-balanced distribution in the downsampled histogram:
in practice with our  dataset, we found a value of $0.5$ to be a good
trade-off.